\documentclass[12pt]{article}
\usepackage{a4wide,amsmath,amssymb}

\newcommand{\cuba}{\textsc{Cuba}}
\newcommand{\ie}{i.e.\ }
\newcommand{\eg}{e.g.\ }
\newcommand{\rd}{\mathrm{d}}
\newcommand{\order}{\mathcal{O}}
\newcommand{\epsabs}{\varepsilon_{\text{abs}}}
\newcommand{\epsrel}{\varepsilon_{\text{rel}}}
\newcommand{\nnew}{n_s^{\text{new}}}
\newcommand{\nmin}{n_s^{\text{min}}}
\newcommand{\nmax}{n_s^{\text{max}}}
\newcommand{\nneed}{n_{\text{need}}}
\newcommand{\tot}{_{\text{tot}}}
\newcommand{\ctot}{_{c,\text{tot}}}
\newcommand{\cvec}[1]{\vec #1}
\newcommand{\dvec}[1]{\mathbf{#1}}
\newcommand{\norm}[1]{\|#1\|_1}
\newcommand{\lbrac}{\symbol{123}}
\newcommand{\rbrac}{\symbol{125}}
\newcommand{\VarIn}[1]{\item\texttt{#1} \textit{$\langle$in\/$\rangle$},}
\newcommand{\VarOut}[1]{\item\texttt{#1} \textit{$\langle$out\,$\rangle$},}
\newcommand{\Option}[2]{\item\texttt{#1 -> #2},}

\makeatletter
\def\reportno#1{\gdef\@reportno{#1}}
\def\@maketitle{%
  \hfill{\small\begin{tabular}[t]{r}%
    \@reportno
  \end{tabular}\par}%
  \vskip 2em%
  \begin{center}%
    \let \footnote \thanks
    {\large \@title \par}%
    \vskip 1.5em%
    {%\large
      \lineskip .5em%
      \begin{tabular}[t]{c}%
        \@author  
      \end{tabular}\par}%
    \vskip 1em%
    {%\large
     \@date}%
  \end{center}%
  \par
  \vskip 1.5em}
\makeatother

\begin{document}

\reportno{MPP--2004--40\\hep--ph/0404043}

\title{\cuba\ -- a library for multidimensional numerical integration}

\author{T. Hahn \\
Max-Planck-Institut f\"ur Physik \\
F\"ohringer Ring 6, D--80805 Munich, Germany}

\date{January 26, 2005}

\maketitle

\begin{abstract}
The \cuba\ library provides new implementations of four general-purpose
multidimensional integration algorithms: Vegas, Suave, Divonne, and
Cuhre.  Suave is a new algorithm, Divonne is a known algorithm to which
important details have been added, and Vegas and Cuhre are new
implementations of existing algorithms with only few improvements over
the original versions.  All four algorithms can integrate vector
integrands and have very similar Fortran, C/C++, and Mathematica
interfaces.
\end{abstract}

%------------------------------------------------------------------------

\section{Introduction}

Many problems in physics (and elsewhere) involve computing an integral,
and often enough this has to be done numerically, as the analytical
result is known only in a limited number of cases.  In one dimension,
the situation is quite satisfactory: standard packages, such as
\textsc{Quadpack} \cite{quadpack}, reliably integrate a broad class of
functions in modest CPU time.  The same is unfortunately not true for
multidimensional integrals.

This paper presents the \cuba\ library with new implementations of four
algorithms for multidimensional numerical integration: Vegas, Suave,
Divonne, and Cuhre.  They have a C/C++, Fortran, and Mathematica
interface each and are invoked in a very similar way, thus making them
easily interchangeable, \eg for comparison purposes.  All four can
integrate vector integrands.  Cuhre is a deterministic algorithm, the
others use Monte Carlo methods.

Vegas is the simplest of the four.  It uses importance sampling for
variance reduction, but is only in some cases competitive in terms of
the number of samples needed to reach a prescribed accuracy. 
Nevertheless, it has a few improvements over the original algorithm
\cite{Vegas1,Vegas2} and comes in handy for cross-checking the results
of other methods.

Suave is a new algorithm which combines the advantages of two popular
methods: importance sampling as done by Vegas and subregion sampling in
a manner similar to Miser \cite{Miser}.  By dividing into subregions,
Suave manages to a certain extent to get around Vegas' difficulty to
adapt its weight function to structures not aligned with the coordinate
axes.

Divonne is a further development of the CERNLIB routine D151
\cite{Divonne}.  Divonne works by stratified sampling, where the
partitioning of the integration region is aided by methods from
numerical optimization.  A number of improvements have been added to
this algorithm, the most significant being the possibility to supply
knowledge about the integrand.  Narrow peaks in particular are difficult
to find without sampling very many points, especially in high
dimensions.  Often the exact or approximate location of such peaks is
known from analytic considerations, however, and with such hints the
desired accuracy can be reached with far fewer points.

Cuhre\footnote{%
	The D from the original name was dropped since the \cuba\ 
	library uses double precision throughout.}
employs a cubature rule for subregion estimation in a globally adaptive
subdivision scheme \cite{dcuhre}.  It is hence a deterministic, not a
Monte Carlo method.  In each iteration, the subregion with the largest
error is halved along the axis where the integrand has the largest
fourth difference.  Cuhre is quite powerful in moderate dimensions, and
is usually the only viable method to obtain high precision, say relative
accuracies much below $10^{-3}$.

The new algorithms were coded from scratch in C, which is a compromise
of sorts between C++ and Fortran 77, combining ease of linking to
Fortran code with the availability of reasonable memory management.  The
declarations have been chosen such that the routines can be called from
Fortran directly.  The Mathematica versions are based on the same C code
and use the MathLink API to communicate with Mathematica.

%------------------------------------------------------------------------

\section{Vegas}

Vegas is a Monte Carlo algorithm that uses importance sampling as a
variance-reduction technique.  Vegas iteratively builds up a
piecewise constant weight function, represented on a rectangular grid. 
Each iteration consists of a sampling step followed by a refinement of
the grid.  The exact details of the algorithm can be found in
\cite{Vegas1,Vegas2} and shall not be reproduced here.

Changes with respect to the original version are:
\begin{itemize}
\item
Sobol quasi-random numbers \cite{Sobol} rather than pseudo-random
numbers are used by default.  Empirically, this seems to accelerate
convergence quite a bit, most noticeably in the early stages of the
integration.

From theoretical considerations it is of course known (see \eg
\cite{Niederreiter}) that quasi-random sequences yield a convergence
rate of $\order(\log^{n_d} n_s/n_s)$, where $n_d$ is the number of
dimensions and $n_s$ the number of samples, which is much better than
the usual $\order(1/\sqrt{n_s})$ for ordinary Monte Carlo.  But these
convergence rates are meaningful only for large $n_s$ and so it came as
a pleasant surprise that the gains are considerable already at the
beginning of the sampling process.  It shows that quasi-Monte Carlo
methods blend well with variance-reduction techniques such as importance
sampling.

Similarly, it was not clear from the outset whether the statistical
standard error would furnish a suitable error estimate since
quasi-random numbers are decidedly non-random in a number of respects. 
Yet also here empirical evidence suggests that the standard error works
just as well as for pseudo-random numbers.

\item
The present implementation allows the number of samples to be increased
in each iteration.  With this one can mimic the strategy of calling
Vegas with a small number of samples first to `get the grid right' and
then using an alternate entry point to perform the `full job' on the
same grid with a larger number of samples.

\item
The option to add simple stratified sampling on top of the importance
sampling, as proposed in the appendix of \cite{Vegas1}, has not been
implemented in the present version.  Tests with the Vegas version from
\cite{NumRecipes}, which contains this feature, showed that convergence
was accelerated only when the original pseudo-random numbers were used
and that with quasi-random numbers convergence was in fact even slower 
in some cases.
\end{itemize}
Vegas' major weakness is that it uses a separable (product) weight
function.  As a consequence, Vegas can offer significant improvements
only as far as the integrand's characteristic regions are aligned with
the coordinate axes.

%------------------------------------------------------------------------

\section{Suave}

Suave (short for \textsc{su}bregion-\textsc{a}daptive \textsc{ve}gas)
uses Vegas-like importance sampling combined with a globally adaptive
subdivision strategy: Until the requested accuracy is reached, the
region with the largest error at the time is bisected in the dimension
in which the fluctuations of the integrand are reduced most.  The number
of new samples in each half is prorated for the fluctuation in that
half.

A similar method, known as recursive stratified sampling, is implemented
in Miser \cite{Miser}.  Miser always samples a fixed number of points,
however, which is somewhat undesirable since it does not stop once the
prescribed accuracy is reached.

Suave first samples the integration region in a Vegas-like step, \ie
using importance sampling with a separable weight function.  It then
slices the integration region in two, as Miser would do.  Suave does not
immediately recurse on those subregions, however, but maintains a list
of all subregions and selects the region with the largest absolute error
for the next cycle of sampling and subdivision.  That is, Suave uses
global error estimation and terminates when the requested relative or
absolute accuracy is attained.

The information on the weight function collected in one Vegas step is
not lost.  Rather, the grid from which the weight function is computed
is stretched and re-used on the subregions.  A region which is the
result of $m - 1$ subdivisions thus has had $m$ Vegas iterations
performed on it.

The improvements over Vegas and Miser come at a price, which is the
amount of memory required to hold all the samples.  Memory consumption
is not really severe on modern hardware, however.  The component that
scales worst is the one proportional to the number of samples, which is
$$
8 (n_d + n_c + 1) n_s\text{ bytes}\,,
$$
where $n_d$ is the number of dimensions of the integral, $n_c$ the
number of components of the integrand, and $n_s$ the number of samples. 
For a million samples on a scalar integrand of 10 variables, this works
out to 96 megabytes -- not all that enormous these days.

\subsection{Description of the algorithm}

As Suave is a new algorithm, the following description will be fairly
detailed.  For greater notational clarity, $n_c$-dimensional vectors are
denoted with a vector arrow ($\cvec f$\,) and $n_d$-dimensional vectors
with boldface letters ($\dvec x$) in the following, where $n_d$ is the
dimension of the integral and $n_c$ the number of components of the
integrand.

The essential inputs are $\epsrel$ and $\epsabs$, the relative and
absolute accuracies, $\nnew$, the number of samples added in each
iteration, $\nmax$, the maximum number of samples allowed, and $p$, a
flatness parameter described below.

Suave has a main loop which calls a Vegas-like sampling step.  The main
loop is responsible for subdividing the subregions and maintaining the
totals.  The sampling step does the actual sampling on the subregions
and computes the region results.

\subsubsection{Main loop}
\label{sect:suavemain}

\begin{enumerate}
\item
Initialize the random-number generator and allocate a data structure for
the entire integration region.  Initialize its Vegas grid with
equidistant bins.

\item
Sample the entire integration region with $\nnew$ points. This gives an
initial estimate of the integral $\cvec I\tot$, the variance 
$\cvec\sigma\tot^{\,2}$, and $\cvec\chi\tot^{\,2}$.

\item
Find the component $c$ for which $r_c = \sigma\ctot/
\max(\epsabs, \epsrel I\ctot)$ is maximal.

If none of the $r_c$'s exceeds unity, indicate success and return.

\item
If the number of samples spent so far equals or exceeds $\nmax$,
indicate failure and return.

\item
Find the region $r$ with the largest $\sigma_c^2$.

\item
Find the dimension $d$ which minimizes $F_c(r_L^d) + F_c(r_R^d)$, where
$r_{L,R}^d$ are the left and right halves of $r$ with respect to $d$. 
$F_c(r_{L,R}^d)$ is the fluctuation of the samples that fall into
$r_{L,R}^d$ and is computed as
\begin{equation}
\label{eq:fluct}
F_c(r_{L,R}^d)
= \biggl[\left\|
    1 + \tilde F_c(\dvec x_i\in r_{L,R}^d)
  \right\|_p\biggr]^{2/3}
= \biggl[\sum\left|
    1 + \tilde F_c(\dvec x_i\in r_{L,R}^d)
  \right|^p\,\biggr]^{2/(3p)},
\end{equation}
where all samples $\dvec x_i$ that fall into the respective half are 
used in the norm/sum and the single-sample fluctuation $\tilde F_c$ is 
defined as
$$
\tilde F_c(\dvec x) =
  w(\dvec x) \, \left|\frac{f_c(\dvec x) - I_c(r)}{I_c(r)}\right|
             \, \frac{|f_c(\dvec x) - I_c(r)|}{\sigma_c(r)}\,.
$$
This empirical recipe combines the relative deviation from the region
mean, $(f - I)/I$, with the $\chi$ value, $|f - I|/\sigma$, weighted by 
the Vegas weight $w$ corresponding to sample $\dvec x$.  Note that the 
$I_c$ and $\sigma_c$ values of the entire region $r$ are used.

Samples strongly contribute to $F$ the more they lie away from the
predicted mean \emph{and} the more they lie out of the predicted error
band.  Tests have shown that large values of $p$ are beneficial for
`flat' integrands, whereas small values are preferred if the integrand
is `volatile' and has high peaks.  $p$ has thus been dubbed a flatness
parameter.  The effect comes from the fact that with increasing $p$, $F$
becomes more and more dominated by `outliers,' \ie points with a large 
$\tilde F$.

The power 2/3 in Eq.~(\ref{eq:fluct}) is also used in Miser, where it is
motivated as the exponent that gives the best variance reduction
(\cite{NumRecipes}, p.~315).

\item
Refine the grid associated with $r$, \ie incorporate the information
gathered on the integrand in the most recent sample over $r$ into the
weight function.  This is done precisely as in Vegas (see
\cite{Vegas1}), with the extension that if the integrand has more than
one component, the marginal densities are computed not from $f^2$ but
from the weighted sum\footnote{%
	It is fairly obvious that scale-invariant quantities must be 
	used in the sum, otherwise the component with the largest 
	absolute scale would dominate.  It is less clear whether $\eta_0 
	= (\int f_c\,\rd\dvec x)^2 = I\ctot^2$, $\eta_1 = (\int 
	|f_c|\,\rd\dvec x)^2$, or $\eta_2 = \int f_c^2\,\rd\dvec x$ (or 
	any other) make the best weights.  Empirically, $\eta_0$ turns 
	out to be both slightly superior in convergence and easier to 
	compute than $\eta_1$ and $\eta_2$ and has thus been chosen in 
	Suave.

	A possible explanation for this is that in cases where there 
	are large compensations within the integral, \ie when $\int 
	f_c\,\rd\dvec x\ll\int |f_c|\,\rd\dvec x$, it is particularly 
	necessary for the overall accuracy that component $c$ be sampled 
	accurately, and thus be given more weight in $\overline{f^2}$, 
	and this is better accomplished by dividing $f_c^2$ by the 
	``small'' number  $\eta_0$ than by the ``large'' number $\eta_1$ 
	or $\eta_2$.}
$$
\overline{f^2} = \sum_{c = 1}^{n_c} \frac{f_c^2}{I\ctot^2}\,.
$$

\item
Bisect $r$ in dimension $d$:

Allocate a new region, $r_L$, and copy to $r_L$ those of $r$'s samples
falling into the left half.  Compute the Vegas grid for $r_L$ by
appropriately ``stretching'' $r$'s grid, \ie by interpolating all grid
points of $r$ with values less than 1/2.

Set up $r_R$ for the right half analogously.

\item
Sample $r_L$ with $n_L = \max\Bigl(\frac{F_c(r_L)}{F_c(r_L) + F_c(r_R)}
\nnew, \nmin\Bigr)$ and $r_R$ with $n_R = \max(\nnew - n_L, \nmin)$ 
points, where $\nmin = 10$.

\item
To safeguard against underestimated errors, supplement the variances 
by the difference of the integral values in the following way:
$$
\sigma_{c,\text{new}}^2(r_{R,L}) = \sigma_c^2(r_{R,L})
  \left(1 + \frac{\Delta_c}
                 {\sqrt{\sigma_c^2(r_L) + \sigma_c^2(r_R)}}\right)^2 +   
  \Delta_c^2
$$
for each component $c$, where $\cvec\Delta = \frac 14 |\cvec I(r_L) + 
\cvec I(r_R) - \cvec I(r)|$.

This acts as a penalty for regions whose integral value changes
significantly by the subdivision and effectively moves them up in the
order of regions to be subdivided next.

\item
Update the totals: Subtract $r$'s integral, variance, and $\chi^2$-value
from the totals and add those of $r_L$ and $r_R$.

\item
Discard $r$, put $r_L$ and $r_R$ in the list of regions.

\item
Go to Step 3.
\end{enumerate}

\subsubsection{Sampling step}

The function which does the actual sampling is a modified Vegas
iteration.  It is invoked with two arguments: $r$, the region to be
sampled and $n_m$, the number of new samples.

\begin{enumerate}
\item
Sample a set of $n_m$ new points using the weight function given by the
grid associated with $r$.  For a region which is the result of $m - 1$
subdivisions, the list of samples now consists of $m$ sets of samples.

\item
For each set of samples, compute the mean $\cvec I_i$ and variance
$\cvec\sigma_i^{\,2}$.

\item
Compute the results for the region as
$$
I_c = \frac{\sum_{i = 1}^m w_{i,c} I_{i,c}}{\sum_{i = 1}^m w_{i,c}},
\quad
\sigma_c^2 = \frac 1{\sum_{i = 1}^m w_{i,c}},
\quad
\chi_c^2 = \frac 1{\sigma_c^2}\left[
  \frac{\sum_{i = 1}^m w_{i,c} I_{i,c}^2}{\sum_{i = 1}^m w_{i,c}} -
  I_c^2\right],
$$
where the inverse of the set variances are used as weights, $w_{i,c} =
1/\sigma_{i,c}^2$.  This is simply Gaussian error propagation.

For greater numerical stability, $\chi_c^2$ is actually computed as
$$
\chi_c^2
= \sum_{i = 1}^m w_{i,c} I_{i,c}^2 -
  I_c \sum_{i = 1}^m w_{i,c} I_{i,c}
= \sum_{i = 2}^m w_{i,c} I_{i,c} (I_{i,c} - I_{1,c}) -
  I_c \sum_{i = 2}^m w_{i,c} (I_{i,c} - I_{1,c})\,.
$$
\end{enumerate}

%------------------------------------------------------------------------

\section{Divonne}

Divonne uses stratified sampling for variance reduction, that is, it 
partitions the integration region such that all subregions have an 
approximately equal value of a quantity called the spread $\cvec s$,
defined as
\begin{equation}
\cvec s(r) = \frac 12 V(r)
  \Bigl(\max_{\dvec x\in r} \cvec f(\dvec x) - 
        \min_{\dvec x\in r} \cvec f(\dvec x)\Bigr)\,,
\end{equation}
where $V(r)$ is the volume of region $r$.  What sets Divonne apart from
Suave is that the minimum and maximum of the integrand are sought using
methods from numerical optimization.  Particularly in high dimensions, 
the chance that one of the previously sampled points lies in or even 
close to the true extremum is fairly small.

On the other hand, the numerical minimization is beset with the usual
pitfalls, \ie starting from the lowest of a (relatively small) number of
sampled points, Divonne will move directly into the local minimum
closest to the starting point, which may or may not be close to the
absolute minimum.

Divonne is a lot more complex than Suave and Vegas but also
significantly faster for many integrands.  For details on the methods
used in Divonne please consult the original references \cite{Divonne}. 
New features with respect to the CERNLIB version (Divonne 4) are:
\begin{itemize}
\item
Integration is possible in dimensions 2 through 33 (not 9 as before).  
Going to higher dimensions is a matter of extending internal tables 
only.

\item
The possibility has been added to specify the location of possible
peaks, if such are known from analytical considerations.  The idea here
is to help the integrator find the extrema of the integrand, and narrow
peaks in particular are a challenge for the algorithm.  Even if only the
approximate location is known, this feature of hinting the integrator
can easily cut an order of magnitude out of the number of samples needed
to reach the required accuracy for complicated integrands.  The points
can be specified either statically, by passing a list of points at the
invocation, or dynamically, through a subroutine called for each
subregion.

\item
Often the integrand subroutine cannot sample points lying on or very
close to the integration border.  This can be a problem with Divonne
which actively searches for the extrema of the integrand and homes in on
peaks regardless of whether they lie on the border.  The user may 
however specify a border region in which integrand values are not 
obtained directly, but extrapolated from two points inside the `safe' 
interior.

\item
The present algorithm works in three phases, not two as before.  Phase 1
performs the partitioning as outlined above.  From the preliminary
results obtained in this phase, Divonne estimates the number of samples
necessary to reach the desired accuracy in phase 2, the final
integration phase.  Once the phase-2 sample for a particular subregion
is in, a $\chi^2$ test is used to assess whether the two sample averages
are consistent with each other within their error bounds.  Subregions
which fail this test move on to phase 3, the refinement phase, where
they can be subdivided again or sampled a third time with more points,
depending on the parameters set by the user.

\item
For all three phases the user has a selection of methods to obtain the
integral estimate: a Korobov \cite{Korobov} or Sobol \cite{Sobol}
quasi-random sample of given size, a Mersenne Twister
\cite{MersenneTwister} pseudo-random sample of given size, and the
cubature rules of Genz and Malik \cite{GenzMalik} of degree 7, 9, 11,
and 13 that are also used in Cuhre.  The latter are embedded rules and
hence provide an intrinsic error estimate (that is, an error estimate
not based on the spread).  When this independent error estimate is
available, it supersedes the spread-based error when computing the total
error.  Also, regions whose spread-based error exceeds the intrinsic
error are selected for refinement, too.

In spite of these novel options, the cubature rules of the original 
Divonne algorithm were not implemented.
\end{itemize}

Due to its complexity, the new Divonne implementation was painstakingly
tested against the CERNLIB routine to make sure it produces the same
results before adding the new features.

%------------------------------------------------------------------------

\section{Cuhre}

Cuhre is a deterministic algorithm which uses one of several cubature
rules of polynomial degree in a globally adaptive subdivision scheme.
The subdivision algorithm is similar to Suave's (see Sect.\
\ref{sect:suavemain}) and works as follows:

While the total estimated error exceeds the requested bounds:

1) choose the region with the largest estimated error,

2) bisect this region along the axis with the largest fourth 
   difference,

3) apply the cubature rule to the two subregions,

4) merge the subregions into the list of regions and update the 
   totals.

Details on the algorithm and on the cubature rules employed in Cuhre can
be found in the original references \cite{dcuhre}.  The present
implementation offers only superficial improvements, such as an
interface consistent with the other \cuba\ routines and a slightly
simpler invocation, \eg one does not have to allocate a workspace.

In moderate dimensions Cuhre is very competitive, particularly if the 
integrand is well approximated by polynomials.  As the dimension 
increases, the number of points sampled by the cubature rules rises 
considerably, however, and by the same token the usefulness declines.
For the lower dimensions, the actual number of points that are spent per 
invocation of the basic integration rule are listed in the following 
table.
\begin{center}
\begin{tabular}{l|ccccccccc}
number of dimensions &
	4 & 5 & 6 & 7 & 8 & 9 & 10 & 11 & 12 \\ \hline
points in degree-7 rule &
	65 & 103 & 161 & 255 & 417 & 711 & 1265 & 2335 & 4433 \\
points in degree-9 rule &
	153 & 273 & 453 & 717 & 1105 & 1689 & 2605 & 4117 & 6745
\end{tabular}
\end{center}

%------------------------------------------------------------------------

\section{Download and Compilation}

The source code of the \cuba\ library can be downloaded as a gzipped tar
file from the Web site \texttt{http://www.feynarts.de/cuba}.  The
archive unpacks into a directory \texttt{Cuba-1.1}.  Change into this
directory and type ``make'' to build the library \texttt{libcuba.a} and
the MathLink executables \texttt{Vegas}, \texttt{Suave},
\texttt{Divonne}, and \texttt{Cuhre}.  If Mathematica and/or the mcc
MathLink compiler are not available, type ``make lib'' to build just the
library.

The distribution contains two demonstration programs in Fortran 77 and 
C, as well as the test suite used in Sect.\ \ref{sect:tests}, which is
written in Mathematica.

The code is C99 compliant and compiles flawlessly with the GNU C
compiler, versions 2.95 and higher.  Other C compilers may have
difficulties with inline functions and variable-size arrays, which are
C99 extensions.  In a pinch, edit the makefile and uncomment the line
\begin{verbatim}
   CFLAGS += -DNDIM=8 -DNCOMP=2
\end{verbatim}
This fixes the size of all internal arrays at compile time but, of
course, at most 8-dimensional integrals of at most 2-component
integrands can now be integrated.

Linking Fortran or C/C++ code that uses one of the algorithms is
straightforward, just add \texttt{-lcuba} (for the \cuba\ library) and
\texttt{-lm} (for the math library) to the compiler command line, as in
\begin{verbatim}
   f77 -o myexecutable mysource.f -lcuba -lm
   cc -o myexecutable mysource.c -lcuba -lm
\end{verbatim}

%------------------------------------------------------------------------

\section{User Manual}

\subsection{Usage in Fortran}

Although written in C, the declarations have been chosen such that the
routines are directly accessible from Fortran, \ie no wrapper code is
needed.  In fact, Vegas, Suave, Divonne, and Cuhre can be called as if 
they were Fortran subroutines respectively declared as
\begin{verbatim}
        subroutine vegas(ndim, ncomp, integrand,
     &    epsrel, epsabs, flags, mineval, maxeval,
     &    nstart, nincrease,
     &    neval, fail, integral, error, prob)
\end{verbatim}
\begin{verbatim}
        subroutine suave(ndim, ncomp, integrand,
     &    epsrel, epsabs, flags, mineval, maxeval,
     &    nnew, flatness,
     &    nregions, neval, fail, integral, error, prob)
\end{verbatim}
\begin{verbatim}
        subroutine divonne(ndim, ncomp, integrand,
     &    epsrel, epsabs, flags, mineval, maxeval,
     &    key1, key2, key3, maxpass,
     &    border, maxchisq, mindeviation,
     &    ngiven, ldxgiven, xgiven, nextra, peakfinder,
     &    nregions, neval, fail, integral, error, prob)
\end{verbatim}
\begin{verbatim}
        subroutine cuhre(ndim, ncomp, integrand,
     &    epsrel, epsabs, flags, mineval, maxeval,
     &    key,
     &    nregions, neval, fail, integral, error, prob)
\end{verbatim}

\subsubsection{Common Arguments}

\begin{itemize}
\VarIn{integer ndim}
the number of dimensions of the integral.

\VarIn{integer ncomp}
the number of components of the integrand.

\VarIn{external integrand}
the integrand.  The external subroutine which computes the 
integrand is expected to be declared as
\begin{verbatim}
  subroutine integrand(ndim, x, ncomp, f)
  integer ndim, ncomp
  double precision x(ndim), f(ncomp)
\end{verbatim}

\VarIn{double precision epsrel, epsabs}
the requested relative and absolute accuracies.
The integrator tries to find an estimate $\hat I$ for the integral $I$
which for every component $c$ fulfills $|\hat I_c - I_c|\leqslant
\max(\epsabs, \epsrel I_c)$.

\VarIn{integer flags}
flags governing the integration:
\begin{itemize}
\item Bits 0 and 1 encode the verbosity level, \ie 0 to 3.

Level 0 does not print any output, level 1 prints `reasonable'
information on the progress of the integration, level 2 also echoes the
input parameters, and level 3 further prints the subregion results (if
applicable).

\item Bit 2 = 0,
all sets of samples collected on a subregion during the various 
iterations or phases contribute to the final result.

Bit 2 = 1,
only the last (largest) set of samples is used in the final result.

\item Bit 3 = 0,
Sobol quasi-random numbers are used for sampling,

Bit 3 = 1,
Mersenne Twister pseudo-random numbers are used for sampling.
\end{itemize}
To select \eg Sobol quasi-random numbers, last samples only, and
verbosity level 2, pass 6 = 0 + 4 + 2 for the flags.  The higher bits 
are presently ignored, but should be zero for future compatibility.

\VarIn{integer mineval}
the minimum number of integrand evaluations required.

\VarIn{integer maxeval}
the (approximate) maximum number of integrand evaluations allowed.

\VarOut{integer nregions}
the actual number of subregions needed (not present in Vegas).

\VarOut{integer neval}
the actual number of integrand evaluations needed.

\VarOut{integer fail}
an error flag:
\begin{itemize}
\item
$\mathtt{fail} = 0$, the desired accuracy was reached,
\item
$\mathtt{fail} = -1$, dimension out of range,
\item
$\mathtt{fail} > 0$, the accuracy goal was not met within the allowed
maximum number of integrand evaluations.  While Vegas, Suave, and Cuhre
simply return 1, Divonne can estimate the number of points by which
\texttt{maxeval} needs to be increased to reach the desired accuracy and
returns this value.
\end{itemize}

\VarOut{double precision integral(ncomp)}
the integral of \texttt{integrand} over the unit hypercube.

\VarOut{double precision error(ncomp)}
the presumed absolute error of \texttt{integral}.

\VarOut{double precision prob(ncomp)}
the $\chi^2$-probability (not the $\chi^2$-value itself!) that
\texttt{error} is not a reliable estimate of the true integration 
error\footnote{%
	To judge the reliability of the result expressed through
	\texttt{prob}, remember that it is the null hypothesis that is 
	tested by the $\chi^2$ test, which is that \texttt{error} 
	\emph{is} a reliable estimate.  In statistics, the null 
	hypothesis may be rejected only if \texttt{prob} is fairly close 
	to unity, say $\mathtt{prob} > .95$.}.
\end{itemize}

\subsubsection{Vegas-specific Arguments}
\label{sect:vegasargs}

\begin{itemize}
\VarIn{integer nstart}
the number of integrand evaluations per iteration to start with.

\VarIn{integer nincrease}
the increase in the number of integrand evaluations per iteration.
\end{itemize}
Vegas furthermore allows to store internal parameters for use in
subsequent invocations.  There are two possibilities:
\begin{itemize}
\item
It may accelerate convergence to keep the grid accumulated during one 
integration for the next one, if the integrands are reasonably similar 
to each other.  Vegas maintains an internal table with space for ten 
grids for this purpose.  The slot in this grid is specified by the 
variable
\begin{verbatim}
        integer gridno
        common /vegasgridno/ gridno
\end{verbatim}
If a grid number between 1 and 10 is selected, the grid is not discarded
at the end of the integration, but stored in the respective slot of the 
table for a future invocation.  The grid is only re-used if the 
dimension of the subsequent integration is the same as the one it 
originates from.

\item
Vegas can also store its entire internal state (\ie all the
information to resume an interrupted integration) in an external
file.  To this end, a file name has to be specified in the variable
\begin{verbatim}
        character*128 statefile
        common /vegasstate/ statefile
\end{verbatim}
The state file is updated after every iteration.  If, on a subsequent
invocation, Vegas finds a file of the specified name, it loads the
internal state and continues from the point it left off.  Needless to
say, using an existing state file with a different integrand generally
leads to wrong results. Once the integration finishes successfully, \ie
the prescribed accuracy is attained, the state file is removed.

This feature is useful mainly to define `check-points' in long-running 
integrations from which the calculation can be restarted.
\end{itemize}

\subsubsection{Suave-specific Arguments}

\begin{itemize}
\VarIn{integer nnew}
the number of new integrand evaluations in each subdivision.

\VarIn{double precision flatness}
the parameter $p$ in Eq.~(\ref{eq:fluct}), \ie the type of norm used to
compute the fluctuation of a sample.  This determines how prominently
`outliers,' \ie individual samples with a large fluctuation, figure in
the total fluctuation, which in turn determines how a region is split
up.  As suggested by its name, \texttt{flatness} should be chosen large
for `flat' integrands and small for `volatile' integrands with high
peaks.  Note that since \texttt{flatness} appears in the exponent, one
should not use too large values (say, no more than a few hundred) lest
terms be truncated internally to prevent overflow.
\end{itemize}

\subsubsection{Divonne-specific Arguments}
\label{sect:divonneargs}

\begin{itemize}
\VarIn{integer key1}
determines sampling in the partitioning phase:

$\mathtt{key1} = 7, 9, 11, 13$ selects the cubature rule of degree 
\texttt{key1}.  Note that the degree-11 rule is available only in 3
dimensions, the degree-13 rule only in 2 dimensions.

For other values of \texttt{key1}, a quasi-random sample of
$n_1 = |\mathtt{key1}|$ points is used, where the sign of \texttt{key1}
determines the type of sample,
\begin{itemize}
\item
$\mathtt{key1} > 0$, use a Korobov quasi-random sample,
\item
$\mathtt{key1} < 0$, use a ``standard'' sample
(a Mersenne Twister pseudo-random sample if bit 3 of the
\texttt{flags} is set, otherwise a Sobol quasi-random sample).
\end{itemize}

\VarIn{integer key2}
determines sampling in the final integration phase:

$\mathtt{key2} = 7, 9, 11, 13$ selects the cubature rule of degree 
\texttt{key2}.  Note that the degree-11 rule is available only in 3
dimensions, the degree-13 rule only in 2 dimensions.

For other values of \texttt{key2}, a quasi-random sample is used, where 
the sign of \texttt{key2} determines the type of sample,
\begin{itemize}
\item
$\mathtt{key2} > 0$, use a Korobov quasi-random sample,
\item
$\mathtt{key2} < 0$, use a ``standard'' sample (see description of 
\texttt{key1} above),
\end{itemize}
and $n_2 = |\mathtt{key2}|$ determines the number of points,
\begin{itemize}
\item
$n_2\geqslant 40$, sample $n_2$ points,
\item
$n_2 < 40$, sample $n_2\,\nneed$ points, where $\nneed$ is the number of
points needed to reach the prescribed accuracy, as estimated by Divonne 
from the results of the partitioning phase.
\end{itemize}

\VarIn{integer key3}
sets the strategy for the refinement phase:

$\mathtt{key3} = 0$, do not treat the subregion any further.

$\mathtt{key3} = 1$, split the subregion up once more.

Otherwise, the subregion is sampled a third time with \texttt{key3}
specifying the sampling parameters exactly as \texttt{key2} above.

\VarIn{integer maxpass}
controls the thoroughness of the partitioning phase:
The partitioning phase terminates when the estimated total number of 
integrand evaluations (partitioning plus final integration) does not 
decrease for \texttt{maxpass} successive iterations.

A decrease in points generally indicates that Divonne discovered new
structures of the integrand and was able to find a more effective
partitioning.  \texttt{maxpass} can be understood as the number of
`safety' iterations that are performed before the partition is accepted
as final and counting consequently restarts at zero whenever new
structures are found.

\VarIn{double precision border}
the width of the border of the integration region.  Points falling into
this border region will not be sampled directly, but will be
extrapolated from two samples from the interior.  Use a nonzero 
\texttt{border} if the integrand subroutine cannot produce values
directly on the integration boundary.

\VarIn{double precision maxchisq}
the maximum $\chi^2$ value a single subregion is allowed to have in the
final integration phase.  Regions which fail this $\chi^2$ test and
whose sample averages differ by more than \texttt{mindeviation} move on
to the refinement phase.

\VarIn{double precision mindeviation}
a bound, given as the fraction of the requested error of the entire
integral, which determines whether it is worthwhile further examining a
region that failed the $\chi^2$ test.  Only if the two sampling averages
obtained for the region differ by more than this bound is the region
further treated.

\VarIn{integer ngiven}
the number of points in the \texttt{xgiven} array.

\VarIn{integer ldxgiven}
the leading dimension of \texttt{xgiven}, \ie the offset between one 
point and the next in memory.

\VarIn{double precision xgiven(ldxgiven,ngiven)}
a list of points where the integrand might have peaks.  Divonne will
consider these points when partitioning the integration region.  The
idea here is to help the integrator find the extrema of the integrand in
the presence of very narrow peaks.  Even if only the approximate
location of such peaks is known, this can considerably speed up
convergence.

\VarIn{integer nextra}
the maximum number of extra points the peak-finder subroutine will
return.  If \texttt{nextra} is zero, \texttt{peakfinder} is not called
and an arbitrary object may be passed in its place, \eg just 0.

\VarIn{external peakfinder}
the peak-finder subroutine.  This subroutine is called whenever a region 
is up for subdivision and is supposed to point out possible peaks lying 
in the region, thus acting as the dynamic counterpart of the static list 
of points supplied in \texttt{xgiven}.  It is expected to be declared as
\begin{verbatim}
  subroutine peakfinder(ndim, b, n, x)
  integer ndim, n
  double precision b(2,ndim)
  double precision x(ldxgiven,n)
\end{verbatim}
The bounds of the subregion are passed in the array \texttt{b}, where 
\texttt{b(1,$d$)} is the lower and \texttt{b(2,$d$)} the upper bound in 
dimension $d$.  On entry, \texttt{n} specifies the maximum number of 
points that may be written to \texttt{x}.  On exit, \texttt{n} must 
contain the actual number of points in \texttt{x}.
\end{itemize}
In contrast to the other algorithms, Divonne passes the integrand one 
more argument, \ie the integrand subroutine is really declared as
\begin{verbatim}
  subroutine integrand(ndim, x, ncomp, f, phase)
  integer ndim, ncomp, phase
  double precision x(ndim), f(ncomp)
\end{verbatim}
The fifth argument, \texttt{phase}, indicates the integration phase:
\begin{itemize}
\item 0, sampling of the points in \texttt{xgiven},
\item 1, partitioning phase,
\item 2, final integration phase,
\item 3, refinement phase.
\end{itemize}
This information might be useful if the integrand takes long to compute
and a sufficiently accurate approximation of the integrand is available. 
The actual value of the integral is only of minor importance in the
partitioning phase, which is instead much more dependent on the peak
structure of the integrand to find an appropriate tessellation.  An
approximation which reproduces the peak structure while leaving out the
fine details might hence be a perfectly viable and much faster
substitute when \texttt{phase\,.lt.\,2}.

In all other instances, \texttt{phase} can be ignored and it is
entirely admissible to declare the integrand with only four arguments.

\subsubsection{Cuhre-specific Arguments}

\begin{itemize}
\VarIn{integer key}
chooses the basic integration rule:

$\mathtt{key} = 7, 9, 11, 13$ selects the cubature rule of degree 
\texttt{key}.  Note that the degree-11 rule is available only in 3
dimensions, the degree-13 rule only in 2 dimensions.

For other values, the default rule is taken, which is the degree-13 rule 
in 2 dimensions, the degree-11 rule in 3 dimensions, and the degree-9 
rule otherwise.
\end{itemize}

\subsection{Usage in C/C++}

Being written in C, the algorithms can of course be used in C/C++ 
directly.  The declarations are as follows:
\begin{verbatim}
typedef void (*integrand_t)(const int *, const double [],
  const int *, double []);
\end{verbatim}
\begin{verbatim}
void Vegas(const int ndim, const int ncomp, integrand_t integrand,
  const double epsrel, const double epsabs,
  const int flags, const int mineval, const int maxeval,
  const int nstart, const int nincrease,
  int *neval, int *fail,
  double integral[], double error[], double prob[])
\end{verbatim}
\begin{verbatim}
void Suave(const int ndim, const int ncomp, integrand_t integrand,
  const double epsrel, const double epsabs,
  const int flags, const int mineval, const int maxeval,
  const int nnew, const double flatness,
  int *nregions, int *neval, int *fail,
  double integral[], double error[], double prob[])
\end{verbatim}
\begin{verbatim}
void Divonne(const int ndim, const int ncomp, integrand_t integrand,
  const double epsrel, const double epsabs,
  const int flags, const int mineval, const int maxeval,
  const int key1, const int key2, const int key3,
  const int maxpass, const double border,
  const double maxchisq, const double mindeviation,
  const int ngiven, const int ldxgiven, double xgiven[],
  const int nextra,
  void (*peakfinder)(const int *, const double [], int *, double []),
  int *nregions, int *neval, int *fail,
  double integral[], double error[], double prob[])
\end{verbatim}
\begin{verbatim}
void Cuhre(const int ndim, const int ncomp, integrand_t integrand,
  const double epsrel, const double epsabs,
  const int flags, const int mineval, const int maxeval,
  const int key,
  int *nregions, int *neval, int *fail,
  double integral[], double error[], double prob[])
\end{verbatim}
These prototypes are contained in \texttt{cuba.h} which should (in C) or
must (in C++) be included when using the \cuba\ routines.  The arguments
are as in the Fortran case, with the obvious translations, \eg
\texttt{double precision} = \texttt{double}.  Note, however, the
declarations of the integrand and peak-finder functions, which expect
pointers to integers rather than integers.  This is required for
compatibility with Fortran.

For convenience, the \texttt{Divonne} prototype glosses over the fact
that Divonne passes an optional fifth argument to the integrand (see end
of Sect.\ \ref{sect:divonneargs}).  Usually the integrand is declared
with only four arguments since this extra information is not needed. 
With the `correct' prototype, the compiler would only generate
unnecessary warnings (in C) or errors (in C++).  In the rare cases where
the integrand really has five arguments, an explicit typecast to
\verb=integrand_t= must be used in the invocation of Divonne.

The global variables for the grid number and state file used in Vegas 
(see Sect.\ \ref{sect:vegasargs}) are also defined in \texttt{cuba.h} as
\begin{verbatim}
   extern int vegasgridno_;
   extern char vegasstate_[128];
\end{verbatim}

\subsection{Usage in Mathematica}

The Mathematica versions are based on essentially the same C code and
communicate with Mathematica via the MathLink API.  When building the
package, the executables \texttt{Vegas}, \texttt{Suave},
\texttt{Divonne}, and \texttt{Cuhre} are compiled for use in
Mathematica.  In Mathematica one first needs to load them with the 
\texttt{Install} function, as in
\begin{verbatim}
   Install["Divonne"]
\end{verbatim}
which makes a Mathematica function of the same name available.  These 
functions are used almost like \texttt{NIntegrate}, only some options 
are different.  For example,
\begin{verbatim}
   Vegas[x^2/(Cos[x + y + 1] + 5), {x,0,5}, {y,0,5}]
\end{verbatim}
integrates a scalar function, or
\begin{verbatim}
   Suave[{Sin[z] Exp[-x^2 - y^2],
          Cos[z] Exp[-x^2 - y^2]}, {x,-1,1}, {y,-1,3}, {z,0,1}]
\end{verbatim}
integrates a vector.  As is evident, the integration region can be 
chosen different from the unit hypercube.  Innermore boundaries may 
depend on outermore integration variables, \eg
\verb=Cuhre[1, {x,0,1}, {y,0,x}]= gives the area of the unit triangle.

The functions return a list which contains the results for each
component of the integrand in a sublist \{integral estimate, estimated 
absolute error, $\chi^2$ probability\}.  For the Suave example above 
this would be
\begin{verbatim}
   {{1.1216, 0.000991577, 0.0000104605}, 
    {2.05246, 0.00146661, 0.00920716}}
\end{verbatim}
The other parameters are specified via the following options.  Default 
values are given on the right-hand sides of the rules.

\subsubsection{Common Options}

\begin{itemize}
\Option{PrecisionGoal}{3}
the number of digits of relative accuracy to seek, that is, $\epsrel =
10^{-\mathtt{PrecisionGoal}}$.

\Option{AccuracyGoal}{12}
the number of digits of absolute accuracy to seek, that is, $\epsabs =
10^{-\mathtt{AccuracyGoal}}$.
The integrator tries to find an estimate $\hat I$ for the integral $I$
which for every component $c$ fulfills $|\hat I_c - I_c|\leqslant
\max(\epsabs, \epsrel I_c)$.

\Option{MinPoints}{0}
the minimum number of integrand evaluations required.

\Option{MaxPoints}{50000}
the (approximate) maximum number of integrand evaluations allowed.

\Option{Verbose}{1}
how much information to print on intermediate results, can take values
from 0 to 3.

Level 0 does not print any output, level 1 prints `reasonable'
information on the progress of the integration, level 2 also echoes the
input parameters, and level 3 further prints the subregion results (if
applicable).  Note that the subregion boundaries in the level-3 printout
refer to the unit hypercube, \ie are possibly scaled with respect to the
integration limits passed to Mathematica.  This is because the
underlying C code, which emits the output, is unaware of any scaling of
the integration region, which is done entirely in Mathematica.

\Option{Final}{All}
whether only the last (largest) or all sets of samples collected on a
subregion during the various iterations or phases contribute to the
final result.

\Option{PseudoRandom}{False}
whether Mersenne Twister pseudo-random numbers are used for sampling
instead of Sobol quasi-random numbers.

\Option{Regions}{False}
whether to return the tessellation of the integration region (thus not
present in Vegas, which does not partition the integration region).

If \texttt{Regions -> True} is chosen, a two-component list is returned,
where the first element is the list of regions, and the second element
is the integration result as described above.  Each region is specified
in the form \texttt{Region[$x_{\mathrm{ll}}$,\,$x_{\mathrm{ur}}$,\,%
$\mathit{res}$,\,$\mathit{df}$]}, where $x_{\text{ll}}$ and
$x_{\text{ur}}$ are the multidimensional equivalents of the lower left
and upper right corner, \textit{res} is the integration result for the
subregion, given in the same form as the total result but with the
$\chi^2$ value instead of the $\chi^2$ probability, and \textit{df} are
the degrees of freedom corresponding to the $\chi^2$ values.  

Cuhre cannot state a $\chi^2$ value separately for each region, hence
the $\chi^2$ values and degrees of freedom are omitted from the
\texttt{Region} information.

\Option{Compiled}{True}
whether to compile the integrand function before use.  Note two caveats:
\begin{itemize}
\item
The function values still have to pass through the MathLink interface,
and in the course of this are truncated to machine precision.  Not
compiling the integrand will thus in general not deliver more accurate
results.
\item
Compilation should be switched off if the compiled integrand shows
unexpected behaviour.  As the Mathematica online help points out, ``the
number of times and the order in which objects are evaluated by
\texttt{Compile} may be different from ordinary Mathematica code.''
\end{itemize}
\end{itemize}

\subsubsection{Vegas-specific Options}

\begin{itemize}
\Option{NStart}{1000}
the number of integrand evaluations per iteration to start with.

\Option{NIncrease}{500}
the increase in the number of integrand evaluations per iteration.

\Option{GridNo}{0}
the slot in the internal grid table.

It may accelerate convergence to keep the grid accumulated during one
integration for the next one, if the integrands are reasonably similar
to each other.  Vegas maintains an internal table with space for ten
grids for this purpose.  If a \texttt{GridNo} between 1 and 10 is
chosen, the grid is not discarded at the end of the integration, but
stored for a future invocation.  The grid is only re-used if the
dimension of the subsequent integration is the same as the one it
originates from.

\Option{StateFile}{""}
the file name for storing the internal state.  If a non-empty string is
given here, Vegas will store its entire internal state (\ie all the
information to resume an interrupted integration) in this file after
every iteration.  If, on a subsequent invocation, Vegas finds a file of
the specified name, it loads the internal state and continues from the
point it left off.  Needless to say, using an existing state file with a
different integrand generally leads to wrong results.  Once the
integration finishes successfully, \ie the prescribed accuracy is
attained, the state file is removed.

This feature is useful mainly to define `check-points' in long-running 
integrations from which the calculation can be restarted.
\end{itemize}

\subsubsection{Suave-specific Options}

\begin{itemize}
\Option{NNew}{1000}
the number of new integrand evaluations in each subdivision.

\Option{Flatness}{50}
the parameter $p$ in Eq.~(\ref{eq:fluct}), \ie the type of norm used to
compute the fluctuation of a sample.  This determines how prominently
`outliers,' \ie individual samples with a large fluctuation, figure in
the total fluctuation, which in turn determines how a region is split
up.  As suggested by its name, \texttt{Flatness} should be chosen large
for `flat' integrands and small for `volatile' integrands with high
peaks.  Note that since \texttt{Flatness} appears in the exponent, one
should not use too large values (say, no more than a few hundred) lest
terms be truncated internally to prevent overflow.
\end{itemize}

\subsubsection{Divonne-specific Options}

\begin{itemize}
\Option{Key1}{47}
an integer which governs sampling in the partitioning phase:

$\mathtt{Key1} = 7, 9, 11, 13$ selects the cubature rule of degree 
\texttt{Key1}.  Note that the degree-11 rule is available only in 3
dimensions, the degree-13 rule only in 2 dimensions.

For other values of \texttt{Key1}, a quasi-random sample of
$n_1 = |\mathtt{Key1}|$ points is used, where the sign of \texttt{Key1}
determines the type of sample,
\begin{itemize}
\item
$\mathtt{Key1} > 0$, use a Korobov quasi-random sample,
\item
$\mathtt{Key1} < 0$, use a ``standard'' sample
(a Mersenne Twister pseudo-random sample for
\texttt{PseudoRandom -> True}, otherwise a Sobol quasi-random sample).
\end{itemize}

\Option{Key2}{1}
an integer which governs sampling in the final integration phase:

$\mathtt{Key2} = 7, 9, 11, 13$ selects the cubature rule of degree 
\texttt{Key2}.  Note that the degree-11 rule is available only in 3
dimensions, the degree-13 rule only in 2 dimensions.

For other values of \texttt{Key2}, a quasi-random sample is used, where 
the sign of \texttt{Key2} determines the type of sample,
\begin{itemize}
\item
$\mathtt{Key2} > 0$, use a Korobov quasi-random sample,
\item
$\mathtt{Key2} < 0$, use a ``standard'' sample
(see description of \texttt{Key1} above),
\end{itemize}
and $n_2 = |\mathtt{Key2}|$ determines the number of points,
\begin{itemize}
\item
$n_2\geqslant 40$, sample $n_2$ points,
\item
$n_2 < 40$, sample $n_2\,\nneed$ points, where $\nneed$ is the number of
points needed to reach the prescribed accuracy, as estimated by Divonne 
from the results of the partitioning phase.
\end{itemize}

\Option{Key3}{1}
an integer which sets the strategy for the refinement phase:

$\mathtt{Key3} = 0$, do not treat the subregion any further.

$\mathtt{Key3} = 1$, split the subregion up once more.

Otherwise, the subregion is sampled a third time with \texttt{Key3}
specifying the sampling parameters exactly as \texttt{Key2} above.

\Option{MaxPass}{5}
the number of passes after which the partitioning phase terminates.
The partitioning phase terminates when the estimated total number of 
integrand evaluations (partitioning plus final integration) does not 
decrease for \texttt{MaxPass} successive iterations.

A decrease in points generally indicates that Divonne discovered new
structures of the integrand and was able to find a more effective
partitioning.  \texttt{MaxPass} can be understood as the number of
`safety' iterations that are performed before the partition is accepted
as final and counting consequently restarts at zero whenever new
structures are found.

\Option{Border}{0}
the width of the border of the integration region.  Points falling into
this border region are not sampled directly, but are extrapolated from
two samples from the interior.  Use a nonzero \texttt{Border} if the
integrand function cannot produce values directly on the integration
boundary.

The border width always refers to the unit hypercube, \ie it is not
rescaled if the integration region is not the unit hypercube.

\Option{MaxChisq}{10}
the maximum $\chi^2$ value a single subregion is allowed to have in the
final integration phase.  Regions which fail this $\chi^2$ test and
whose sample averages differ by more than \texttt{MinDeviation} move on
to the refinement phase.

\Option{MinDeviation}{.25}
a bound, given as the fraction of the requested error of the entire
integral, which determines whether it is worthwhile further examining a
region that failed the $\chi^2$ test.  Only if the two sampling averages
obtained for the region differ by more than this bound is the region
further treated.

\Option{Given}{\lbrac\rbrac}
a list of points where the integrand might have peaks.  A point is a
list of $n_d$ real numbers, where $n_d$ is the dimension of the
integral.

Divonne will consider these points when partitioning the integration
region.  The idea here is to help the integrator find the extrema of the
integrand in the presence of very narrow peaks.  Even if only the
approximate location of such peaks is known, this can considerably speed
up convergence.

\Option{NExtra}{0}
the maximum number of points that will be considered in the output of 
the \texttt{PeakFinder} function.

\Option{PeakFinder}{(\lbrac\rbrac\&)}
the peak-finder function.  This function is called whenever a region is
up for subdivision and is supposed to point out possible peaks lying in
the region, thus acting as the dynamic counterpart of the static list of
points supplied with \texttt{Given}.  It is invoked with two arguments,
the multidimensional equivalents of the lower left and upper right
corners of the region being investigated, and must return a (possibly
empty) list of points.  A point is a list of $n_d$ real numbers, where 
$n_d$ is the dimension of the integral.
\end{itemize}

\subsubsection{Cuhre-specific Options}

\begin{itemize}
\Option{Key}{0}
chooses the basic integration rule:

$\mathtt{Key} = 7, 9, 11, 13$ selects the cubature rule of degree 
\texttt{Key}.  Note that the degree-11 rule is available only in 3
dimensions, the degree-13 rule only in 2 dimensions.

For other values, the default rule is taken, which is the degree-13 rule 
in 2 dimensions, the degree-11 rule in 3 dimensions, and the degree-9 
rule otherwise.
\end{itemize}

%------------------------------------------------------------------------

\section{Tests and Comparisons}
\label{sect:tests}

Four integration routines may seem three too many, but as the following 
tests show, all have their strengths and weaknesses.  Fine-tuning the 
algorithm parameters can also significantly affect performance.

In the following, the test suite of Genz \cite{Genz} is used.  Rather
than testing individual integrands, Genz proposes the following six
families of integrands:
\begin{equation}
\label{eq:families}
\begin{array}{ll}
\text{1. Oscillatory:} &
  f_1(\dvec x) = \cos(\dvec c\cdot\dvec x + 2\pi w_1)\,, \\[2ex]
\text{2. Product peak:} &
  f_2(\dvec x) = \prod\limits_{i = 1}^{n_d}
    \dfrac 1{(x_i - w_i)^2 + c_i^{-2}}\,, \\[3ex]
\text{3. Corner peak:} &
  f_3(\dvec x) = \dfrac 1{(1 + \dvec c\cdot\dvec x)^{n_d + 1}}\,, \\[3ex]
\text{4. Gaussian:} &
  f_4(\dvec x) = \exp(-\dvec c^2 (\dvec x - \dvec w)^2)\,, \\[2ex]
\text{5. $C^0$-continuous:}\quad &
  f_5(\dvec x) = \exp(-\dvec c\cdot |\dvec x - \dvec w|)\,, \\[2ex]
\text{6. Discontinuous:} &
  f_6(\dvec x) = \begin{cases}
    0 & \text{for }x_1 > w_1 \vee x_2 > w_2\,, \\
    \exp(\dvec c\cdot\dvec x) & \text{otherwise}.
  \end{cases}
\end{array}
\end{equation}
Parameters designated by $w$ are non-affective, they vary \eg the 
location of peaks, but should in principle not affect the difficulty of 
the integral.

Parameters designated by $c$ are affective and in a sense ``define''
the difficulty of the integral, \eg the width of peaks are of this kind. 
The $c_i$ are positive and the difficulty increases with $\norm{\dvec c} 
= \sum_{i = 1}^{n_d} c_i$.

The testing procedure is thus: Choose uniform random numbers from
$[0,1)$ for the $c_i$ and $w_i$.  Renormalize $\dvec c$ for a given
difficulty.  Run the algorithms with the integrands thus determined. 
Repeat this procedure 20 times and take the average.

For comparison, Mathematica's \texttt{NIntegrate} function was included
in the test.  Unfortunately, when a maximum number of samples is
prescribed, \texttt{NIntegrate} invariably uses non-adaptive methods, by
default the Halton--Hammersley--Wozniakowski quasi-Monte Carlo
algorithm.  The comparison may thus seem not quite balanced, but this is
not entirely true: Lacking an upper bound on the number of integrand
evaluations, \texttt{NIntegrate}'s adaptive method in some cases `locks
up' (spends an inordinate amount of time and samples) and the user can
at most abort a running calculation, but not extract a preliminary
result.  The adaptive method could reasonably be used only for some of
the integrand families in the test, and it was felt that such a
selection should not be done, as the comparisons should in the first
place give an idea about the \emph{average} performance of the
integration methods, without any fine-tuning.

Table \ref{tab:comp} gives the results of the tests as described above. 
This comparison chart should be interpreted with care, however, and
serves only as a rough measure of the performance of the integration
methods.  Many integrands appearing in actual calculations bear few or
no similarities with the integrand families tested here, and neither
have the integration parameters been tuned to `get the most' out of each
method.

The Mathematica code of the test suite is included in the downloadable 
\cuba\ package.

\begin{table}
$$
\begin{array}{|c|r@{\,\pm\,}r|r@{\,\pm\,}r|r@{\,\pm\,}r
                |r@{\,\pm\,}r|r@{\,\pm\,}r|}
\multicolumn{11}{c}{n_d = 5} \\ \hline
j &
\multicolumn{2}{|c|}{\text{Vegas}} &
\multicolumn{2}{|c|}{\text{Suave}} &
\multicolumn{2}{|c|}{\text{Divonne}} &
\multicolumn{2}{|c|}{\text{Cuhre}} &
\multicolumn{2}{|c|}{\text{NIntegrate}} \\ \hline
1 & 162000 &     0 &
    127300 & 32371 &
     21313 & 11039 &
       819 &     0 &
    218281 &     0 \\
2 &  11750 &  1795 &
     13500 &  1539 &
     17353 &  3743 &
     56238 & 40917 &
    218281 &     0 \\
3 &  16125 &  2411 &
     11500 &  1000 &
     17208 &  2517 &
      1174 &   444 &
    218281 &     0 \\
4 &  56975 & 11372 &
     20100 &  4745 &
     19636 &  6159 &
     22577 & 31424 &
    218281 &     0 \\
5 &  14600 &  3085 &
     15250 &  2337 &
     21675 &  4697 &
    150423 &     0 &
    218281 &     0 \\
6 &  19750 &  4999 &
     23850 &  2700 &
     39694 & 14001 &
      1884 &   215 &
    218281 &     0 \\ \hline
\multicolumn{11}{c}{} \\
\multicolumn{11}{c}{n_d = 8} \\ \hline
j &
\multicolumn{2}{|c|}{\text{Vegas}} &
\multicolumn{2}{|c|}{\text{Suave}} &
\multicolumn{2}{|c|}{\text{Divonne}} &
\multicolumn{2}{|c|}{\text{Cuhre}} &
\multicolumn{2}{|c|}{\text{NIntegrate}} \\ \hline
1 & 153325 & 20274 &
    124350 & 35467 &
     28463 & 31646 &
      3315 &     0 &
    212939 & 13557 \\
2 &  12650 &  1987 &
     21050 &  4594 &
     22030 &  3041 &
     91826 & 58513 &
    218281 &     0 \\
3 &  24325 &  3753 &
     29350 &  3588 & 
     67104 & 16906 & 
     18785 & 22354 &
    218281 &     0 \\
4 &  38575 & 16169 & 
     29250 &  8873 &
     24849 &  5015 &
     62322 & 44328 &
    218281 &     0 \\
5 &  15150 &  2616 &
     25500 &  6444 &
     32885 &  5945 &
    151385 &     0 &
    218281 &     0 \\
6 &  18875 &  2512 &
     40900 &  7196 &
    116744 & 32533 &    
      9724 &  9151 &
    218281 &     0 \\ \hline
\multicolumn{11}{c}{} \\
\multicolumn{11}{c}{n_d = 10} \\ \hline
j &
\multicolumn{2}{|c|}{\text{Vegas}} &
\multicolumn{2}{|c|}{\text{Suave}} &
\multicolumn{2}{|c|}{\text{Divonne}} &
\multicolumn{2}{|c|}{\text{Cuhre}} &
\multicolumn{2}{|c|}{\text{NIntegrate}} \\ \hline
1 & 156050 & 21549 & 
    129800 & 30595 & 
     32176 & 30424 & 
      7815 &     0 &
    214596 & 16481 \\
2 &  14175 &  2672 &
     24800 &  5464 &
     25684 &  7582 &
    144056 & 25983 &
    218281 &     0 \\
3 &  30275 &  6296 &
     51150 & 15608 & 
    139737 & 18505 & 
    109150 & 58224 &
    218281 &     0 \\
4 &  29475 & 10277 &
     34050 & 10200 & 
     27385 &  8498 &
    105763 & 49789 & 
    218281 &     0 \\
5 &  16150 &  2791 &
     31400 &  7715 &
     44393 & 18654 & 
    153695 &     0 & 
    218281 &     0 \\
6 &  22100 &  3085 &
     74900 & 32203 &
    136508 & 17067 & 
     73200 & 64621 &
    218281 &     0 \\ \hline
\end{array}
$$
Test parameters:
\begin{itemize}
\item number of dimensions: $n_d = 5, 8, 10$,
\item requested relative accuracy: $\epsrel = 10^{-3}$,
\item maximum number of samples: $\nmax = 150000$,
\item integrand difficulties:
$
\begin{array}{r||c|c|c|c|c|c}
\text{Integrand family }j &  1  &  2   &  3  &  4   &  5   &  6  \\ 
\hline
\norm{\dvec c_j}          & 6.0 & 18.0 & 2.2 & 15.2 & 16.1 & 16.4
\end{array}
$
\end{itemize}

\caption{\label{tab:comp}The number of samples used, averaged from 20
randomly chosen integrands from each integrand family $j$ defined in
Eq.~(\ref{eq:families}).  Values in the vicinity of $\nmax$ generally
indicate failure to converge. \texttt{NIntegrate} seems not to be able
to stop at around the limit of \texttt{MaxPoints -> $\nmax$}, but always
samples considerably more points.}
\end{table}

%------------------------------------------------------------------------

\section{Summary}

The \cuba\ library offers a choice of four independent routines for
multidimensional numerical integration: Vegas, Suave, Divonne, and
Cuhre.  They work by very different methods, summarized in the following
table (MT = Mersenne Twister):
\begin{center}
\begin{small}
\begin{tabular}{llll}
Routine  &
	Basic integration method &
	Algorithm type &
	Variance reduction \\ \hline \\[-1.5ex]
Vegas &
	Sobol quasi-random sample &
	Monte Carlo &
	importance sampling \\
&
	\textit{or} MT pseudo-random sample &
	Monte Carlo \\[1.5ex]
Suave &
	Sobol quasi-random sample &
	Monte Carlo &
	globally adaptive subdivision \\
&
	\textit{or} MT pseudo-random sample &
	Monte Carlo \\[1.5ex]
Divonne &
	Korobov quasi-random sample &
	Monte Carlo &
	stratified sampling, \\
&
	\textit{or} Sobol quasi-random sample &
	Monte Carlo &
	\quad aided by methods from \\
&
	\textit{or} MT pseudo-random sample &
	Monte Carlo &
	\quad numerical optimization \\
&
	\textit{or} cubature rules &
	deterministic \\[1.5ex]
Cuhre &
	cubature rules &
	deterministic &
	globally adaptive subdivision
\end{tabular}
\end{small}
\end{center}

All four have a C/C++, Fortran, and Mathematica interface and can
integrate vector integrands.  Their invocation is very similar, so it is
easy to substitute one method by another for cross-checking.  For
further safeguarding, the output is supplemented by a $\chi^2$
probability which quantifies the reliability of the error estimate.

The source code is available from \texttt{http://www.feynarts.de/cuba}
and compiles with gcc, the GNU C compiler.  The C functions can be
called from Fortran directly, so there is no need for adapter code. 
Similarly, linking Fortran code with the library is straightforward
and requires no extra tools.

The routines in the \cuba\ library have all been carefully tested, but
it would of course be folly to believe they are completely error-free. 
The author welcomes any kind of feedback, in particular bug and 
performance reports, at hahn@feynarts.de.

\section*{Acknowledgements}

I thank A.~Hoang for involving me in a discussion out of which the 
concept of the Mathematica interface was born and T.~Fritzsche, 
M.~Rauch, and A.M.~de~la~Ossa for testing.

\end{document}